\documentstyle[12pt,twoside]{article}
\newcommand{\be}{\begin{equation}}
\newcommand{\ee}{\end{equation}}
\newcommand{\bea}{\begin{eqnarray}}
\newcommand{\eea}{\end{eqnarray}}
\begin{document}

\begin{center}
{\Large\bf Brownian Motion: Theory and Experiment \\
A simple classroom measurement of the diffusion coefficient}\\[20mm]

Kasturi Basu \footnote{Kasturi Basu did her graduation in Physics from Jadavpur 
University, Kolkata. She is about 
to join the University of Cambridge as a Chevening scholar for the next two years for getting 
a master's degree in Physics. She has special interest in Condensed Matter Physics. 
Kasturi's e-mail address: kasturi\_basu@yahoo.com} 
and Kopinjol Baishya \footnote{ Kopinjol Baishya is a final year M.Sc. student at Delhi University. After M.Sc. he looks
 forward to a research career in Physics. This is part 
of the project work he and Kasturi did at RRI, Bangalore, under the guidance of Dr. Abhishek 
Dhar, during the Summer Programme organised by the institute in May-June, 2002.
Kopinjol's e-mail address: kopinjol@hotmail.com}\\
\end{center}

\bigskip

\begin{abstract}
Brownian motion is the perpetual irregular motion exhibited by small particles immersed in a fluid. 
Such random motion of the particles is produced by statistical fluctuations in the collisions they 
suffer with the molecules of the surrounding fluid. Brownian motion of particles in a fluid (like 
milk particles in water) can be observed under a microscope.
Here we describe a simple experimental set-up to observe Brownian motion and a method of determining the diffusion
coefficient of the Brownian particles,  
based on a theory due to Smoluchowski. While looking through the microscope we focus attention on a 
fixed small volume, and record the number of particles that are trapped in that volume, at regular intervals 
of time. This gives us a time-series 
data, which is enough to determine the diffusion coefficient of the particles to a good degree of 
accuracy. 
\end{abstract}

\bigskip

\section{Description of the experiment}

As our system we took water containing a small number of milk particles. From our experience we found that 
freshly boiled and cooled milk, free from cream works well. A dilution of approximately 
1 drop of milk in 10 drops of water was used since this was found to be suitable for our observations.
Observations were made with a  microscope (the one we used had 15x eye-piece and 45x 
objective). The solution was placed between two pieces of
transparencies (sealed off at the edges to prevent convection currents
which, if not driven off, can drive things crazy) [See Fig.1]. One
could see Brownian motion of the milk particles if one looked at them
carefully  for long enough. However  the motion is not as vigorous or
spectacular as we are used to seeing in textbook pictures.

If we focus attention on a single particle, it is expected that the particle will trace a random path [as in Fig. 2].
Supposing that in time $t$ the particle is displaced from the origin to the position ${\bf r}$. Then, for a Brownian 
particle the mean square displacement is given by $< {\bf r}^2 > =2 D t$ and this defines the diffusion constant of
 the particle in the given fluid. A straightforward way of measuring the diffusion constant $D$ would thus be to
 make a  large number of observations on the displacement of the particle in a given time $t$ and find the mean 
square displacement. In our experiment however $D$ turns out to be
very small
and so this method is not particularly easy. Hence we use a different method in which we estimate the diffusion constant in a rather indirect manner.
We shall describe the theory behind this method in successive sections.  

We counted the number of particles contained in a well-defined square area in the 
field of view of the microscope at constant intervals of time. We did
it $30$ seconds apart and made $130$ observations. For 
this purpose a fine grid was made on a small piece of transparency
sheet, by etching with a sharp edge, and attached  
to the eyepiece. The area actually enclosed by the grid was calibrated  with 
a standard grid ( we took one that is used in pathological
laboratories for counting blood corpuscles). The time-series data
(time versus number of particles) thus obtained has information on
both static properties, such as the distribution of number of
particles in the observed area, and also dynamic properties such as
the diffusion constant. We first look at the static properties  and
then go on to the problem of extracting the diffusion constant.   

\section{Number distribution: The Poisson distribution}

We plotted the 
distribution of the frequency $W(n)$ with which different numbers of particles $n$ are observed 
within the marked region and compared it with the Poisson distribution that we expect on 
the basis of the theory  which we shall briefly describe
now. 
We consider a well-defined small element of volume $v$ in a much larger volume $V$ containing a large 
number of Brownian particles under conditions of diffusion equilibrium [See fig.3]. The number of 
particles contained in $v$ fluctuates with time.

Now, if we count the number of particles contained in $v$ at constant intervals of time $\tau$
apart, then the frequency $W(n)$ with which different numbers $n$ of particles are counted will 
follow the Poisson distribution:
\be
W(n) = \frac{e^{-\nu} \nu^n}{n!}
\ee
where $\nu$ denotes the average number of particles contained in $v$.

That the distribution will indeed have the form (1) can be proven, provided the following 
assumptions are held true in the given system:

~~~~(i) the particles do not interfere with each other, and 

~~~~(ii) they are equally likely to be anywhere within the system.

Surely, for our system of Brownian particles these two assumptions are really good ones.

In Fig. 4 we show comparison of our experimental data with the Poisson
Distribution, and find that they match pretty well, so as to justify the
assumptions above.

We would like to discuss something more before we come down to tackling the problem at hand, 
that of determining the diffusion coefficient from the time-series data obtained from observation.

\section{The Brownian particle as a random walker}

Each Brownian particle in a fluid is being kicked by fluid molecules randomly, i.e. from all 
sides with equal probability. A Brownian particle can, hence, satisfactorily be modeled as a 
random walker, i.e. a walker who moves in successive steps, each step being independent of 
the previous step.

In the experiment we perform, we are essentially looking at Brownian motion in 2-dimensions  
(with the field of view of the microscope constituting the xy-plane). So now we consider a  
random walker in 2-dimensions. Let's look at the discrete case first. The random walker is 
moving on a 2-dimensional lattice, and can move up, down, right or left with equal probabilities. 
We assume steps of equal size $a$, the time for each step being $\tau$.
Then if at some time $t$ the walker is at the point ${\bf r} = (x , y)$, then at time $t + \tau$ 
it can be at either of the four points $(x+a , y), (x-a , y), (x , y+a), (x , y-a)$. The 
probability of it being at any of these four points is $1/4$. 
Let $Q({\bf r}, t)$ be the probability for the walker to be at the point ${\bf r}$ at time $t$.
Then, 
\bea
Q({\bf r}, t) = \frac 1 4 [ Q(x+a , y , t-\tau) + Q(x-a , y , t-\tau) + Q(x , y+a , t-\tau) + 
Q(x , y-a , t-\tau) ]. \nonumber
\eea
Subtracting $Q({\bf r} , t-\tau)$ from both sides, dividing by $\tau$, and defining $D = a^2/4\tau$, 
we get 
\bea
\frac{Q({\bf r},t) - Q({\bf r},t-\tau)}{\tau} &=& \frac{a^2}{4\tau}~ [ \frac{ Q(x+a,y,t-\tau)
+ Q(x-a,y,t-\tau) -2 Q(x,y,t-\tau)}{a^2} \nonumber\\
&+& \frac{Q(x,y+a,t-\tau) + Q(x,y-a,t-\tau) - 2 Q(x,y,t-\tau)}{a^2} ]\nonumber
\eea
Or, in the continuum limit,
\be
\frac{\partial Q({\bf r},t)}{\partial t} = D~\left[ \frac{\partial^2 Q({\bf r},t)}{\partial x^2}
+ \frac{\partial^2 Q({\bf r},t)}{\partial y^2} \right]
\ee
which is the 2-dimensional diffusion equation, $D$ being the diffusion coefficient.
Solving this ( for initial conditions $Q({\bf r},t = 0) = \delta ({\bf
r})$, i.e. initially the particle is located at the origin) using Fourier transforms, we get 
\be
Q({\bf r},t) = \frac 1 {4\pi D t}~ e^{- \frac{r^2}{4 D t}}.
\ee
Thus, coming back to our system we conclude that the probability of occurrence of a Brownian 
particle at ${\bf r_2}$ at time $t$ when it was at ${\bf r_1}$ at time $t = 0$ is given by
\be
Q({\bf r_1} ; {\bf r_2}) = \frac 1 {4\pi D t} \exp\left[ \frac{-|{\bf r_2} - {\bf r_1}|^2}{4 D t} 
\right].
\ee

\section{Extraction of diffusion coefficient from the time-series data}

We begin by defining the {\it probability $P$ that a particle somewhere inside $v$ will have come 
out of it during the time $\tau$}. (N.B. As mentioned in section 3. we are into a 2-dimensional 
problem. Here we are no longer considering $v$ as the 3-dimensional small volume, but rather 
as its projection onto the $xy$-plane. The symbol $v$ mentioned henceforth is the 2-dimensional area we 
are focusing on and taking snapshots of).

From Eq.(4) and according to the above definition of $P$
\be
P = \frac 1 {4 \pi D \tau v}~ \int_{{\bf r_1} \in v} \exp \left[ \frac{-|{\bf r_2} - {\bf r_1}|^2}{4 D \tau} 
\right] d {\bf r_1} d {\bf r_2}
\ee
where the integration over ${\bf r_1}$ is extended over all points in the interior of $v$ 
while that over ${\bf r_2}$ is extended over all points exterior to $v$.

Alternatively, we can also write 
\be
1 - P = \frac 1 {4 \pi D \tau v}~ \int_{{\bf r_1, r_2} \in v} \exp \left[ \frac{-|{\bf r_2} - 
{\bf r_1}|^2}{4 D \tau} \right] d {\bf r_1} d {\bf r_2}
\ee
where, now, the integration over {\it both} ${\bf r_1}$ and ${\bf r_2}$ are extended over all points 
{\it inside} $v$. (Physically, $1 - P$ is the probability that a particle somewhere inside $v$ will 
have still remained within $v$ after a time $\tau$.)

Now, if we take $v$ to be a square area of side $h$ (which can easily be ensured by using a 
square grid), then Eq.(6) leads to 
\bea
1 - P &=& \frac 1 {4 \pi D \tau h^2}~ \int_0^h \int_0^h dx_1 dx_2 \exp [ - (x_2 - x_1)^2/4 D \tau ]~
\int_0^h \int_0^h dy_1 dy_2 \exp [ - (y_2 - y_1)^2/4 D \tau ] \nonumber\\
&=& \left[ \frac 1 {(4\pi D \tau)^{1/2} h} ~\int_0^h \int_0^h dx_1 dx_2 \exp [ - (x_2 - x_1)^2/4 D \tau ]
\right]^2. \nonumber
\eea
Putting in the substitutions   
\be
\xi = \frac x {2 \sqrt{D \tau}}~;~~~~ \alpha = \frac h {2 \sqrt{D \tau}} ,
\ee
we have 
\bea
1 - P = \left[ \frac 1 {\alpha \sqrt{\pi}} \int_0^{\alpha} \int_0^{\alpha} \exp [- (\xi_1 - \xi_2)^2]
d \xi_1 d \xi_2 \right]^2. \nonumber
\eea
With $\xi_1 - \xi_2 = \eta$, we have
\bea
1 - P = \left[ \frac 2 {\alpha \sqrt{\pi}} \int_0^{\alpha} d \xi_1 \int_0^{\xi_1} d\eta e^{-\eta^2}
\right]^2. \nonumber
\eea
After an integration by parts, we finally have
\be
1 - P = \left[ \frac 2 {\sqrt{\pi}} \int_0^{\alpha} d\xi e^{-\xi^2} - \frac 1 {\alpha \sqrt{\pi}}
[1 - e^{- \alpha^2}] \right]^2.
\ee
The $\alpha$ in Eq.(8) contains $D$, vide Eq.(7).
Thus Eq.(8) relates $P$ with $D$.

The next question, which arises naturally, is how to find $P$ from the experimental time-series 
data. We now show how this can be obtained from the mean square
fluctuation in the difference in particle number in successive counts, which we will
denote by $\langle~\Delta^2~\rangle_{av}$. We will need
to use the Poisson distribution discussed in section 1.  

\section{Determining $P$ experimentally}

Let $\tau$ be the experimental time interval between successive observations.
Let $x$ be the number of particles that remain in the volume $v$ after time $\tau$, having started with $n$ particles within $v$.

Let $y$ be the number of particles that enter the volume $v$ in time $\tau$. Obviously $y$ is independent of $n$, the initial number of particles within $v$, vide assumption (i).

We define the probability that $i$ particles leave volume $v$ in time $\tau$, having started with $n$ particles as $A_i^{(n)}$, the probability that $x$ particles remain in volume $v$ after time $\tau$, starting with $n$ particles within $v$ as $B_x^{(n)}$, and the probability that $y$ particles enter volume $v$ in time $\tau$ as $E_y$.

We have, with reference to the probability $P$ defined at the beginning of section 4,
\be
A_i^{(n)} = \frac{n!}{i!(n - i)!}~P^i~(1 - P)^{n - i},
\ee
which is a Bernoulli distribution.

Now, our system is under equilibrium conditions and there is no net inward or outward flux of particles from volume $v$, i.e. the a priori  probabilities for the entrance and emergence of particles from the volume $v$ are equal.
So $E_y$ will be the same as the probability that $y$ particles emerge from the volume $v$ in time $\tau$ in general, i.e. starting with any value of $n$ (the distribution of $n$ satisfies the Poisson distribution (1) ).
i.e.,
\be
E_y = \sum_{n = y}^\infty~W(n)~A_y^{(n)}.
\ee
From eqs. (1), (9) and (10), we eventually have
\be
E_y = \frac{e^{-\nu~P}~(\nu~P)^y}{y!}.
\ee
which is again a Poisson distribution.
Also, we have, similar to (9),
\be 
B_x^{(n)} = \frac{n!}{x!(n - x)!}~P^{n-x}~(1 - P)^x.
\ee

From equations (12) and (11) we may calculate the mean and mean square deviations of $x$ and $y$, which turn out as
\bea
\langle~x~\rangle_{av} &=& n~(1-P)\nonumber \\
\langle~(~x - \langle~x~\rangle_{av})^2~\rangle_{av} &=& nP(1 - P)\nonumber \\
\langle~y~\rangle_{av} &=& \nu~P\nonumber \\
\langle~(~y - \langle~y~\rangle_{av})^2~\rangle_{av} &=& \nu~P.
\eea
Now, let us consider a situation in which we count $n$ particles within the volume $v$ in one snapshot and then find $m$ particles in $v$ in the next snapshot (i.e., after time $\tau$).
If $T(n,m)$ is the probability of such a transition, evidently,
\be
T(n , m) = \sum_{x + y = m}~B_x^{(n)}~E_y.
\ee
It can be proven, given eq. (14), that the mean and mean square deviations for the distribution of $m$ are respectively the sums of those of $x$ and $y$,
i.e.,
\bea
\langle~m~\rangle_{av} &=& \langle~x~\rangle_{av} + \langle~y~\rangle_{av}~;
\nonumber\\
\langle~(~m - \langle~m~\rangle_{av})^2~\rangle_{av} &=& \langle~(~x - \langle~x~\rangle_{av})^2~\rangle_{av} + \langle~(~y - \langle~y~\rangle_{av})^2~\rangle_{av}.
\nonumber
\eea
Hence we have
\bea
\langle~m~\rangle_{av} = n(1 - P) + \nu~P~;\nonumber \\
\langle~(~m - \langle~m~\rangle_{av})^2~\rangle_{av} = nP(1 - P) + \nu~P.
\eea
We define a quantity $\Delta_n = m - n$. Averaging $\Delta_n^2$ over
all possible values of $m$ (keeping $n$ fixed) using eq. (15), we get
\be
\langle~\Delta_n^2~\rangle_{av} = P^2 [~(\nu - n)^2 - n~] + (n + \nu)~P.
\ee
Now, we further average $\langle~\Delta_n^2~\rangle_{av}$ over all values of $n$ with the weight function $W(n)$ given by eq. (1).
We have,
\bea
\langle~\Delta^2~\rangle_{av} &=& \langle~\langle~\Delta_n^2~\rangle_{av}~\rangle_{av}\nonumber \\
&=& 2\nu~P.
\eea
We note that $\langle~\Delta^2~\rangle_{av}$ is just the mean square
fluctation in the difference in particle number in successive counts,
and $\nu$ is the mean number of particles. These are quantities  
that may easily be calculated from the experimental time-series
data. Thus eq. (17) tells us how to find $P$ experimentally. Finally, 
using eqs. (8)  we can determine the diffusion coefficient from the
experimental time series data.

\section{Experimental results}
For our experiment, $P$ turned out as $0.326$, and the diffusion coefficient $D$ as $2.1 \times 10^{-12} m^2.s^{-1}$.
This is almost ten times greater than the rough theoretical estimate
using Einstein's formula [Ref. Box 1].
Given the simplicity and lack of sophistication of the experimental setup, this is
a reasonable estimate. For obtaining better accuracy, a camera could
be fitted to the eye-piece of the microscope in order to take
snapshots at regular intervals instead of the manual procedure we have
used. This would lead to more accurate data collection and also
much more number of data points (we took only $130$ data points). A
vibration-free and tilt-free working table could also be used to
reduce external disturbances. Finally the milk particles used in the
experiment were not monodisperse, that is there was quite some
variation in the sizes of particles. Using monodisperse particles
would clearly be an improvement.

\section{Suggested Reading}

[1] S.Chandrasekhar, Stochastic Problems in physics and Astronomy, Reviews of Modern Physics, Vol. 15, No. 1, p.44-52,1943

Finally, the authors would like to acknowledge the enlightening discussions on the entire matter that they had with Abhishek Dhar, Supurna Sinha, Joseph Samuel and Sarasij Ray Chaudhuri at RRI.

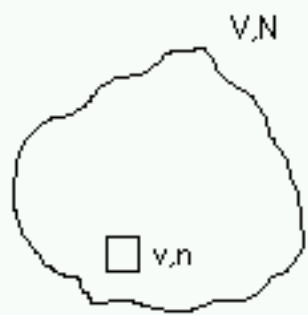
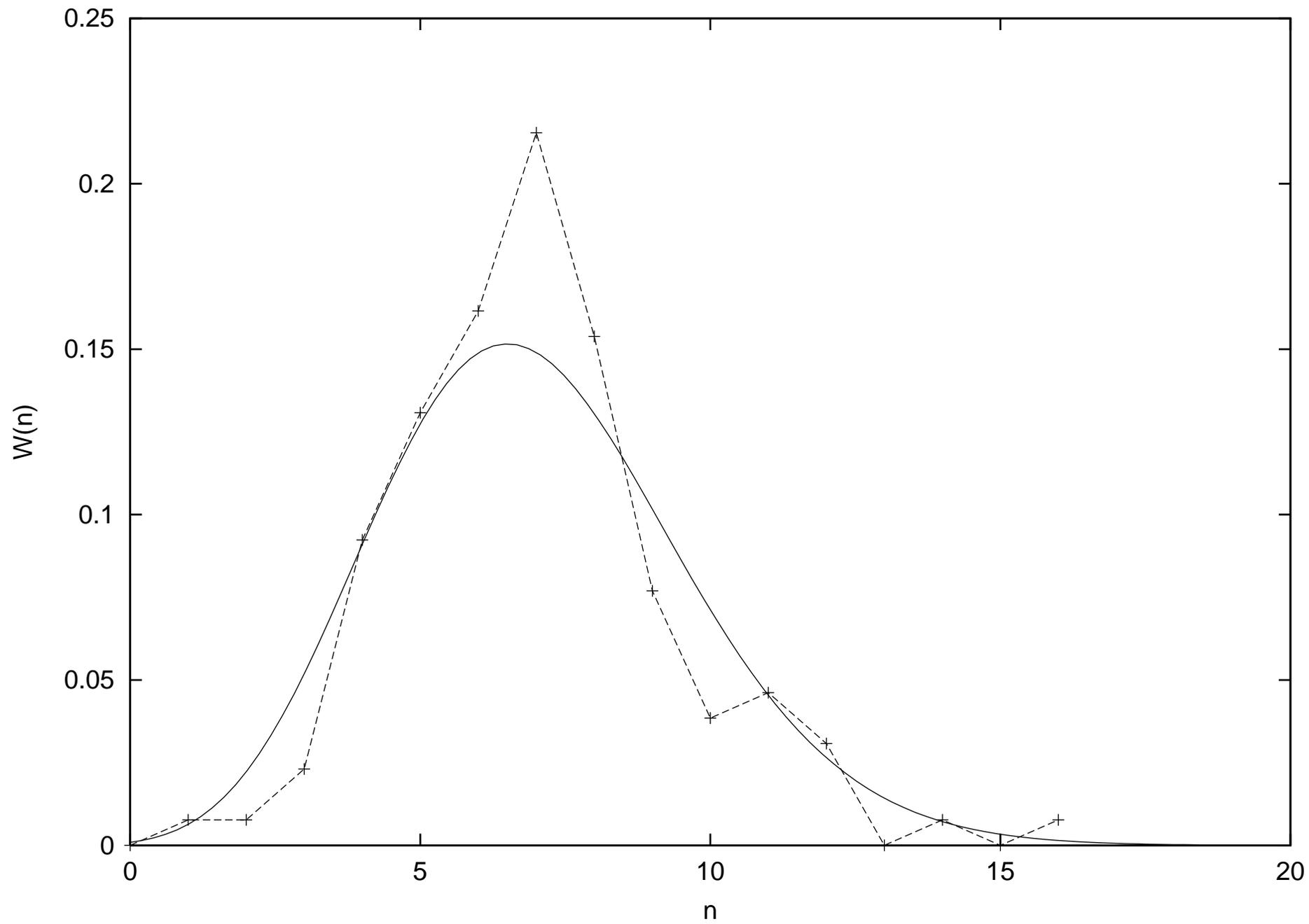
\end{document}